# Constructing Basis Path Set by Eliminating Path Dependency

Juanping Zhu[1*] Qi Meng[2] Wei Chen[2] Yue Wang[3] Zhi-ming Ma[4]
1. Yunnan University, 2. Microsoft Research Asia, 3. Beijing Jiaotong University, 4. Chinese Academy of Science
* Corresponding author

**Abstract:** The way the basis path set works in neural network remains mysterious, and the generalization of newly appeared $\mathcal{G}$-SGD algorithm to more practical network is hindered. The *Basis Path Set Searching* problem is formulated from the perspective of graph theory, to find the basis path set in a regular complicated neural network. Our paper aims to discover the underlying cause of path dependency between two independent substructures. Algorithm DEAH is designed to solve the *Basis Path Set Searching* problem by eliminating such path dependency. The path subdivision chain is proposed to effectively eliminate the path dependency inside the chain and between chains. The theoretical proofs and analysis of polynomial time complexity are presented. The paper therefore provides one methodology to find the basis path set in a more general neural network, which offers theoretical and algorithmic support for the application of $\mathcal{G}$-SGD algorithm in more practical scenarios.

# 1 Introduction

Neural network with ReLU activation function has been developed for a variety of tasks, such as distribution estimation in statistics, machine translation and language modeling etc. [1,2]. How to train the weights of ReLU neural network in appropriate space affects the network performance and efficiency. In order to handle the mismatch during optimizing neural networks in positively scale-invariant space [3,4], Meng et al. [5] proposed to optimize the values of basis path set in $\mathcal{G}$-space of ReLU neural networks by stochastic gradient descent algorithm (SGD) [6]. The experiments turned out that this novel $\mathcal{G}$-SGD algorithm [5] with attractive low dimensionality of basis path set significantly outperformed conventional SGD algorithm. In addition, the performance superiority was approved recently in language modeling and machine translation by adopting the concept of basis path set in transformer network [2]. It is promising for $\mathcal{G}$-SGD algorithm to be generalized to more practical neural networks and be applied widely in applications.

In recent years many efforts have been invested into explaining and understanding the overwhelming success of neural network learning methods [7-11]. Despite the experimental success, $\mathcal{G}$-SGD algorithm needs further theoretical investigation on how the inner mechanism of basis paths works in neural networks. For instance, how to analyze the structure relationship between the basis path set and the network and how does the different network structure affect the algorithm searching for basis path set? On the other hand, the generalization of $\mathcal{G}$-SGD algorithm is currently hindered for the implementation of $\mathcal{G}$-SGD algorithm [5] depends on the structure of neural network, because currently $\mathcal{G}$-SGD algorithm can only heuristically find the basis path set in the simple fully connected network, which demands the width of all layer is the same and there is no edge-skipping over layers. However, there are varieties of structures in practical network such as ResNet and DenseNet with different combination of edge-skipping over layers and ununiform layer width.

We resort to graph theory for theoretical support to understand basis path set like the way explaining neural network in group operations and functional perspectives [12,13]. In the perspective of graph theory, Zhu et al.[14] defined basis path and proposed one hierarchical algorithm to find basis path set in each independent substructure. This hierarchical algorithm [14] can handle the network with ununiform layer width and edge-skipping but requires that there exist no shared layers between any two independent substructure paths, which is strict for practical network.

This paper considers the graph-theory-based *Basis Path Set Searching* problem how to find maximal independent path set in the graph of regular fully connected neural network with any ununiform layer width and any edge-skipping without restriction. We first investigate the combinatoric possibility that maximal independent substructures will bring up path dependency. Then the paper proposes dependency eliminating algorithm based on hierarchical idea (DEAH) to solve *Basis Path Set Searching* problem. The hierarchical idea is employed to decompose the complicated neural network into maximally independent substructures and find basis path set for each independent substructure in parallel. Then we need to eliminate the paths which would cause the path dependency when we combine all these basis path sets together. To avoid the enumeration of all possible substructure path pairs for the shared layers, we take advantage of path subdivision chains in the algorithm designing. Lemmas and theorems are provided to guarantee the algorithm can solve *Basis Path Set Searching* problem in polynomial time.

The contributions of this paper are the following: (i) the paper explains the structure relationship of *Basis Path Set Searching* problem from graph theory perspective and provides one polynomial algorithm; (ii) Algorithm DEAH can help overcome the hurdle of current $\mathcal{G}$-SGD algorithm and generalize $\mathcal{G}$-SGD algorithm further; (iii) this paper provides one methodology to find the basis path set in more general neural network, and it can offer theoretical and algorithmic support for the application of $\mathcal{G}$-SGD algorithm in more practical scenarios.

## 2 Preliminary

Regular fully connected neural network is a $L$-layer multi-layer perceptron with weighted edges that can skip over different layers [14]. We denote $i$-th node in $l$-th layer as $O_i^l$ and the node set of the $l$-th layer as $O^l$. $(O_i^l, O_{i'}^{l+j})$ is denoted as the directed edge from layer $l$ to layer $l+j$, which can skip layer $l+1$ till layer $l+j-1$ as shown in Fig. 1, where $1 \leq j \leq L-l, 1 \leq i \leq |O^l|$ and $1 \leq i' \leq |O^{l+j}|$. Within the same layers, the edges are fully connected. Since graph theory provides one effective platform to investigate the paths in the graph intuitively and theoretically [15-17], the paper would interpret neural network as a triple graph $G = (V, E, w)$, where the finite node set $V = O^0 \cup ... \cup O^l ... \cup O^L$ comprising all nodes in neural network $G$ and finite edge set $E = \{ (O_i^l, O_{i'}^{l+j}) | 0 \leq l \leq L-1 \text{ and } 1 \leq j \leq L-l \}$ consisting of all directed fully connected edges between different layers. $w(O_i^l, O_{i'}^{l+j})$ is the weight of edge $(O_i^l, O_{i'}^{l+j}) \in E$, and $m = |E|$ is the number of edges in graph $G$. Let set $P = \{ (O_{i*}^0, O_{i*}^{1'}, ..., O_{i*}^{j'} ..., O_{i*}^L) | 0 < 1' < 2' ... < j' < L \}$ consist of all paths from the input layer to the output layer in network $G$[15,16], where $O_{i*}^l$ is denoted as some node without specified position in the $l$-th layer for simplicity.

**Definition** 1 (independent path set)[14] Given path set $B \subseteq P$, if there exists one path $p \in B$ and another path $q \in B \setminus \{p\}$ such that we can reach path $p$ from path $q$ through finite steps of path addition and path removal within $B$, we call path set $B$ is dependent. Otherwise, we call path set $B$ is independent.

**Definition 2** (maximal independent path set) [14] A path set $B \subseteq P$ of neural network $G$ is maximally independent, if including any other path $p^* \in P \setminus B$ would make $B \cup p^*$ dependent. Hence, a basis path set $B$ of neural network $G$ is a maximal independent subset of $P$.

The definitions of basic path operations such as path addition and path removal [14] can be found in Appendix A.

**Definition 3** (substructure path)[14] Given fully connected neural network G and induced subgraph $G'$ with $V(G') = (O^0, O^{1'}, ..., O^{i'}, ..., O^L)$, where $1 \leq i' \leq L-1$. Let $P(G')$ be the path set from the input layer to the output layer in $G'$. If all paths in $P(G')$ from the input to the output pass through the same layers homogenously, then any path $p \in P(G')$ can be called as the substructure path of neural network $G$. Substructure path $p$ can express the structure information about sub-graph $G'$ in network $G$.

**Definition 4** (substructure path set) Define one induced sub-graph $G^S = (V^S, E^S)$ of fully connected neural network $G$, which is a simplified network with only one node in each layer, where $V^S = \{$one randomly selected $O_{i*}^l \in O^l | l = 0, ..., L\}$ and $E^S = \{(O_{i*}^l, O_{i*}^k) \in E | O_{i*}^l \in V^S, O_{i*}^k \in V^S, 0 \leq l < k \leq L\}$. By breadth first search, we can get all substructure paths starting from node $O_{i*}^0$ to node $O_{i*}^L$ in $G^S$. Denote this substructure path set as $P^S$, which can represent all substructure information of network $G$.

**Definition 5** (maximal independent substructure path set) Given fully connected neural network $G$ and substructure path set $P^S$. The substructure path set $P_{ind}^S \subset P^S$ is called dependent, if there is one substructure path $p \in P_{ind}^S$ and another path $q \in P_{ind}^S \setminus \{p\}$ such that we can reach path $p$ from path $q$ through finite steps of path addition and path removal within $P_{ind}^S$. Otherwise, we call $P_{ind}^S$ is independent. A substructure path set $P_{ind}^S$ of $P^S$ is maximally independent, if including any other path $p^* \in P^S \setminus P_{ind}^S$ would make $P_{ind}^S \cup p^*$ dependent.

There exists one special structure relationship between two independent substructure paths, *i.e.*, path subdivision, which is the adaptation of graph subdivision [18-20] to the path.

**Definition 6** (path subdivision) Given a substructure path set $P^S$ and one substructure path $p \in P^S$ with one edge $e = (u, v) \in E(p)$. If there exists one substructure path $p' \in P^S$ with sub-path $(u, x_1, x_2, ..., x_k, v)$, sub-path $(u, x_1, x_2, ..., x_k, v)$ is called the edge subdivision of edge $e$. We call $e$ the subdivided edge and $x_1, x_2, ..., x_k$ the subdivision vertices. Furthermore, if $p - e = p' - (u, x_1, x_2, ..., x_k, v)$, we call path $p'$ the path subdivision of path $p$ and path $p$ the subdivided path. Especially, we denote by $p'$ the path obtained from $p$ by subdividing the edges $e_1, ..., e_t, ..., e_{T'}$, where each edge $e_t \in E(p)$ is subdivided once for $t \in \{1, ..., T'\}$.

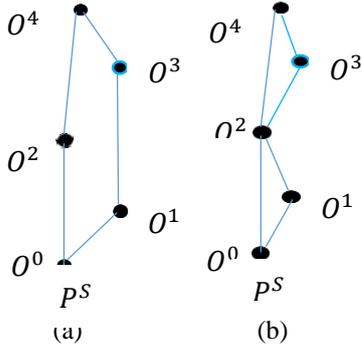 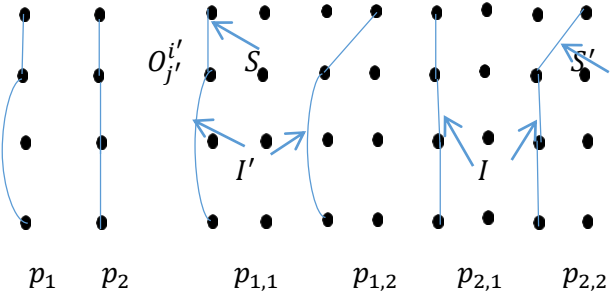

Fig. 1 Example of path subdivision      Fig. 2 Case 1 of path dependency

There is no edge subdivision in $P^S$ in Fig. 1(a). In Fig. 1(b), edge $(O^0, O^2)$ is a subdivided edge, because there exists sub-path $(O^0, O^1, O^2)$ between $O^0$ and $O^2$. Let $p_1 = (O^0, O^1, O^2, O^3, O^4)$, $p_2 = (O^0, O^2, O^3, O^4)$, $p_3 = (O^0, O^1, O^2, O^4)$ and $p_4 = (O^0, O^2, O^4)$. Substructure path $p_1$ is the path subdivision of path $p_2$, $p_3$ and $p_4$. Paths $p_2$ and $p_3$ are the path subdivision of path $p_4$ but $p_2$ and $p_3$ are not path subdivision of each other. Note we use the layer to represent the random node in the layer when discussing the substructure path, which is different from regular path expression.

**Definition 7** (underlying substructure path) Given the substructure path set $P^S$, subset $P' \subset P^S$ and one substructure path $p_0 \in P'$. We call $p_0$ the underlying substructure path of $P'$ if

$p_0$ is the path subdivision of any substructure path $p \in P' \setminus \{p_0\}$.

# 3 Problem Statement

### 3.1 *Basis Path Set Searching* Problem

To solve the challenge related to basis path set in more general neural network mentioned in section Introduction, this paper focuses on *Basis Path Set Searching* Problem, which is defined as follows.

Given graph $G$ representing the regular fully connected network, we aim to find maximal independent path set (basis path set) $B$ in graph $G$.

Here the regular fully connected network can be a network with unbalanced layers (the widths of different layers are not equal) and with edges jumping over different layers.

### 3.2 Path Dependency between Two Independent Substructures

In graph theory, any path $p \in P$ in fully connected network $G$ from the input layer to the output layer can be represented by the basis path set $B$ with smaller cardinality [14]. One hierarchical algorithm has been proposed [14] to decompose the fully connected network into maximal independent substructures $P_{ind}^S$, where each sub-graph $G_r$ induced from $p_r \in P_{ind}^S$ can be treated as a fully connected graph without any edge-skipping. Here $G_r$ is the sub-graph of $G$ by taking all edges from $G$ with the same layers as substructure path $p_r$, in which **Algorithm Subroutine** [14] (in Appendix C) can find basis path set $B_r$ accordingly. It is well-known that hierarchical idea usually decomposes the complicated combinatorial optimization problem into several independent and simpler sub-optimization problems [21-24] and solves each sub-optimization problems separately. Since $P_{ind}^S$ is the maximal independent substructure set and each independent substructure $p_r \in P_{ind}^S$ has unique structure, intuitively we would consider to simply combine these basis path sets $B_r$ together to form basis path set $B$ for network $G$. However, the basis paths from two independent substructures $p_r$ and $p_s$ in our regular graph $G$ couldn't guarantee they are path dependent, though their structures are unique. After investigation, we found the underlying cause of path dependency is $E(p_r) \cap E(p_s) \neq \emptyset$ $(r \neq s)$, *i.e.*, there exist shared edges (layers) between $p_r$ and $p_s$. The shared edges offer the chance for basis paths to exchange the locations of the shared layers between substructures and to cancel same unshared layers within the substructure for the structure uniqueness. There are three typical cases to illustrate this combinatoric possibility, and we also notice that the basis paths from each substructure must appear in pair to cancel the unique unshared layers.

**Case 1**
In Fig. 2, substructure path $p_2$ is the path subdivision of substructure path $p_1$ in the layers of $I$. $p_1$ and $p_2$ share layers of $S$ and $S'$, and $p_1$ and $p_2$ are independent. Given basis paths $p_{1,1}, p_{1,2} \in B_1$, and $p_{2,1}, p_{2,2} \in B_2$. Any path can be represented by the remaining three paths, such as $p_{1,1} = p_{2,1} - p_{2,2} + p_{1,2}$. The same sub-path $I'$ in $p_{1,1}$ and $p_{1,2}$ in the unshared layers can be cancelled inside substructure $p_1$, and the same sub-path $I$ in $p_{2,1}$ and $p_{2,2}$ can be cancelled inside substructure $p_2$ too. In the shared layers, sub-path $S$ of $p_{1,1}$ and $p_{2,1}$ can be swapped, and so does sub-path $S'$ of $p_{1,2}$ and $p_{2,2}$.

**Case 2**
In Fig. 3, both substructure paths $p_2$ and $p_3$ have edge subdivision in each other but they are not path subdivision to each other. $p_2$ and $p_3$ share layers at $S$ and $S'$, so sub-path $S'$ in $p_{2,1}$ and $p_{3,1}$ can be swapped and $S$ in $p_{2,2}$ and $p_{3,2}$ can be swapped. In the unshared layers, sub-path $I' + J'$ in $p_{2,1}$ and $p_{2,2}$ can be cancelled, and $I + J$ in $p_{3,1}$ and $p_{3,2}$ can be cancelled too.

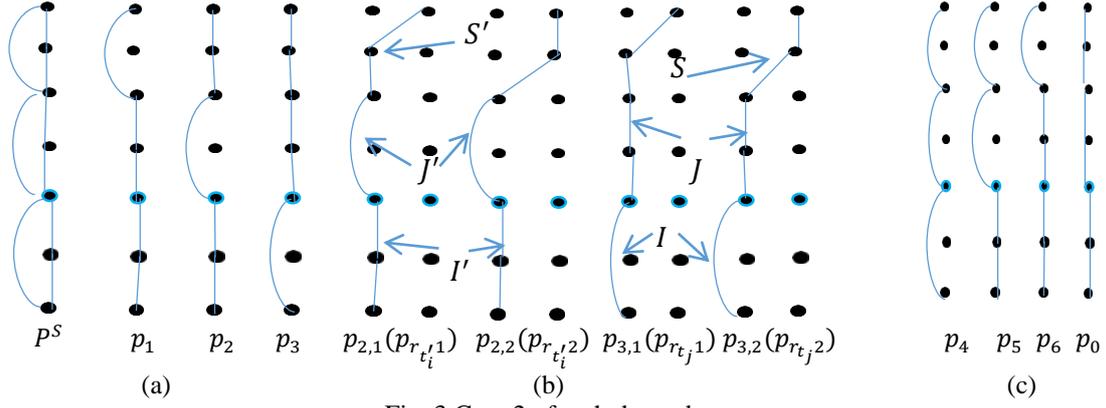

Fig. 3 Case 2 of path dependency

**Case 3**
In Fig. 4, substructure paths $p_2$ and $p_1$ have no edge subdivision in each other, but $p_1$ and $p_2$ have layers of $S$ and $S'$ in common. In the unshared layers, same sub-path $I$ in $p_{1,1}$ and $p_{1,2}$ can be cancelled from each other, and sub-path $I'$ in $p_{2,1}$ and $p_{2,2}$ can be cancelled too. In the shared layers, sub-path $S$ in $p_{1,1}$ and $p_{2,1}$ and $S'$ in $p_{1,2}$ and $p_{2,2}$ can be swapped.

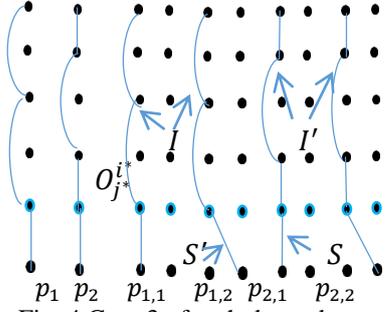

Fig. 4 Case 3 of path dependency

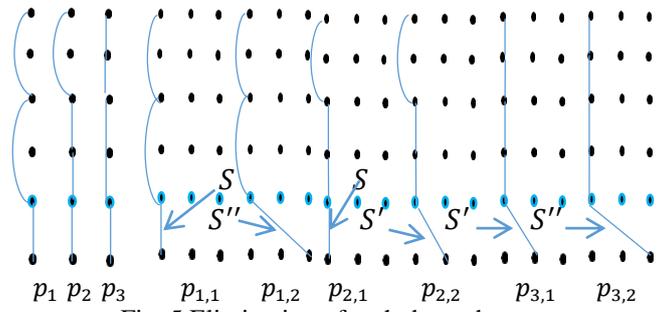

Fig. 5 Elimination of path dependency

**Property 1** If $E(p_r) \cap E(p_s) \neq \emptyset$ $(r \neq s)$, then the path set $B_r \cup B_s$ is not path independent.

**Claim 1** Given two independent substructure paths $p_1$ and $p_2$ with their shared layers $S^* = E(p_1) \cap E(p_2)$, basis paths $p_{1,1}, p_{1,2} \in B_1$ with $S \in E(p_{1,1})$ and $S' \in E(p_{1,2})$ at the shared layers $S^*$. If $p_{1,1} - \sum_{e \in S} e = p_{1,2} - \sum_{e \in S'} e$ at the unshared layers, then there must exist basis paths $p_{2,1}, p_{2,2} \in B_2$ such that $S \in E(p_{2,1})$ and $S' \in E(p_{2,2})$ at the layers of $S^*$ and $p_{2,1} - \sum_{e \in S} e = p_{2,2} - \sum_{e \in S'} e$.
**Proof:** We will prove this claim from two cases.

Case 1. Unshared layers of $I$ appear at the top of shared layers of $S^*$. As shown in Fig. 4, let $I = p_{1,1} - \sum_{e \in S} e = p_{1,2} - \sum_{e \in S'} e$. According to the properties of the direct path [14], the first edge in $I$ must be direct path, because it accepts two different sub-paths $S$ and $S'$ from its incident node, denoted as $O_{j^*}^{i^*}$. Moreover, there must exist one direct path starting from $O_{j^*}^{i^*}$ in $B_2$ regarding $p_2$. Pick up one sub-path starting with this direct path, denoted as $I'$. Here we demand the rule to randomly select paths in **Algorithm Subroutine** be the same while constructing all basis path set. So sub-paths $S$ and $S'$ ending at $O_{j^*}^{i^*}$ must be included in some basis paths of $p_2$ and $I'$ will accept both $S$ and $S'$ in $p_{2,1}$ and $p_{2,2}$ as shown in Fig. 4.

Case 2. If the shared layers of $S^*$ is at the top of unshared layers of $I$ as shown in Fig. 2, $S$ and $S'$ are two different sub-paths starting from $O_{j'}^{i'}$ in $B_1$. According to the same rule to randomly select paths, $S$ and $S'$ must exist as layers of two basis paths in $B_2$. We can pick up the direct path ending at $O_{j'}^{i'}$ which can be concatenated by $S$ and $S'$ at $O_{j'}^{i'}$.

Other locations of shared layers $S^*$ are the combination of Case 1 and Case 2. ∎

**Claim 2** The path dependency couldn't happen among three or more than three independent substructures, if we eliminate the paths which cause path dependency when combing the basis path sets of two independent substructures.

**Proof:** The analysis of path dependency indicates that basis paths from one substructure must appear in pair to cancel unshared layers for the structure uniqueness. Suppose the path dependency happens among three independent substructure paths, *i.e.*, $p_1, p_2$ and $p_3$ as shown in Fig.5. Assume $p_{1,1} - p_{1,2} + p_{2,1} - p_{2,2} = p_{3,1} - p_{3,2}$, where $p_{1,1}, p_{1,2} \in B_1$, $p_{2,1}, p_{2,2} \in B_2$, and $p_{3,1}, p_{3,2} \in B_3$. The common layers $E(S) \subset E(p_{1,1})$ and $E(S) \subset E(p_{2,1})$ must appear in pair for swapping. The same for $E(S') \subset E(p_{2,2})$ and $E(S') \subset E(p_{3,1})$, and $E(S'') \subset E(p_{1,2})$ and $E(S'') \subset E(p_{3,2})$. The unshared layers $p_{1,1} - \sum_{e \in S} e$ and $p_{1,2} - \sum_{e \in S''} e$ must appear the same to cancel the unique structure. So do $p_{2,1} - \sum_{e \in S} e$ and $p_{2,2} - \sum_{e \in S'} e$ pair and $p_{3,1} - \sum_{e \in S'} e$ and $p_{3,2} - \sum_{e \in S''} e$ pair. When we consider eliminating the paths which cause path dependency between two independent substructures, we must delete at least one basis path from corresponding two pairs. For example, if $p_{1,1}$ and $p_{3,2}$ stay, $p_{1,2}$ must be discarded when considering $p_1$ and $p_3$. Therefore, it is impossible to either take swapping for the shared layers or cancel the unshared layers for the basis path from the third substructure such as $p_2$. So, the assumption couldn't happen and Claim 2 holds. ∎

Claim 2 indicates that we only need to consider the path dependency between any two independent substructures. However, it would be expensive if we enumerate all two substructure path pairs with the shared layers in graph $G$. In Case 2 of analysis of path dependency, the maximal independent substructure path set could be $\{p_0, p_1, p_2, p_3\}$ in Fig. 3(a) and it can also be $\{p_0, p_4, p_5, p_6\}$ as in Fig. 3(c), where $p_5$ is the path subdivision of $p_4$, $p_6$ is the path subdivision of $p_5$ and the underlying substructure path $p_0$ is the path subdivision of $p_4, p_5$ and $p_6$. $\{p_0, p_4, p_5, p_6\}$ can form one chain from path subdivision. Motivated by this, path subdivision chain is proposed to avoid complicated enumeration.

**Definition 8** (path subdivision set) Given the substructure path set $P^S$ of fully connected network $G$ and subset $P' \subset P^S$, the path subdivision set $U_r$ of substructure path $p_r \in P'$ is defined as $\{p \in P' | p \text{ is the subdivision of } p_r\}$.

**Definition 9** (path subdivision chain) Given set $\{U_r | r = 1, \ldots, |P'|\}$ of path subdivision sets on $P' \subset P^S$. Based on set containment relationship, we can get $T$ path subdivision chains, *i.e.*, the $t$-th chain $U_{t_1} \supset U_{t_2} \ldots \supset U_{t_j} \ldots \supset U_{t_{s_t}}$, where $\sum_{t=1}^{T} s_t \geq |P'|$ and $U_{t_j} \in \{U_r | r = 1, \ldots, R\}$ is the $j$-th set of the $t$-th chain.

Fig.3(c) can form one path subdivision chain $U_4 \supset U_5 \supset U_6 \supset U_0$ but Fig.3(a) needs 3 chains such as $U_1 \supset U_0, U_2 \supset U_0$ and $U_3 \supset U_0$, though $p_1, p_2$ and $p_3$ have subdivision layers in each other. Obviously, path subdivision set $U_0 = \emptyset$. Some properties and lemmas about the path subdivision chain can be found in Appendix B.

# 4 Dependency Eliminating Algorithm Based on Hierarchical Idea (DEAH)

Algorithm HBPS [14] initiated one inspiring hierarchical idea to decompose the complicated graph $G$ into several independent substructures but the restriction is there doesn't exist shared layers between any two maximal independent substructures, which may cause the path dependency. However, one practical fully connected network allows edge-skipping over different layers and shared layers between different substructures. To solve *Basis Path Set Searching* problem in regular graph $G$, we propose Algorithm DEAH to overcome the restriction of Algorithm HBPS to find basis path set in more practical network by eliminating path dependency.

Step 1 of Algorithm DEAH employs hierarchical idea to decompose the complicated network $G$ into $|P_{ind}^S|$ maximal independent substructures. Compute path subdivision set $U_r$ for $p_r \in P^S$ and sort $\{U_r\}$ with $|U_r|$ in descending order as $\{U_{r_i}|i = 0,1 \ldots, |P^S| - 1\}$ with $U_{r_0} = U_0$. Starting from $i = 0$, get the first $|P_{ind}^S|$ substructure paths from $\{p_{r_i}|i = 0,1 \ldots, |P^S| - 1\}$ such that they are independent. The purpose of this step is to avoid Case 2 in path dependency analysis and stretch the path subdivision chains as long as possible in Step 3. We must pay attention to $|U_{r_{i-1}}| = |U_{r_i}|$ when $Rank(\{U_{r_0}, U_{r_1}, \ldots U_{r_{k'}}, \ldots, U_{r_{i-1}}, U_{r_i}\}) > Rank(\{U_{r_0}, U_{r_1}, \ldots U_{r_{k'}}, \ldots, U_{r_{i-1}}\})$. In this case, if there exists some $U_{r_j} (j < i - 1)$ such that $U_{r_{i-1}} \subset U_{r_j}$ and $U_{r_i} \subset U_{r_j}$ and substructure paths $p_{r_{i-1}}$ and $p_{r_i}$ have edge subdivision in each other, we skip $U_{r_i}$ and go to $U_{r_{i+1}}$. This step rules out the possibility of Case 2 of path dependency (shown in Fig. 3(a)). This is because the chain with $U_{r_{i-1}}$ would split into two chains but we can manipulate to form one chain instead by simply skipping $U_{r_i}$.

Step 2 finds basis path set $B_r$ by calling **Algorithm Subroutine** for each $G_r$ induced by $p_r \in P_{ind}^S$ in parallel, which can be treated as a simple network without edge-skipping. Step 3 is to eliminate the path dependency between different substructure path pairs, by constructing multiple path subdivision chains to avoid enumeration as discussed in Section 3. We scan the selected path subdivision sets in $\{U_r|r = 0,1, \ldots, |P_{ind}^S| - 1\}$ with $|U_r|$ in descending order to stretch the $t$-th chain $U_{t_1} \supset U_{t_2} \ldots \supset U_{t_j} \ldots \supset U_{t_{s_t}} \supset U_0$ as long as possible, and start one new chain starting from $U_r$ if $U_r$ couldn't be contained in some ready chain. Once the multiple chains are established, the next step is to eliminate the path dependency. Algorithm DEAH divides the selected path subdivision sets into two groups. The first group includes the sets which are not the first set in any chain. The second group consists of the first set in all chain. For each set $U_{t_{j+1}}$ in the first group in $t$-th chain, call Algorithm **SDVChain** (in Appendix C) to discard the paths from path set $B'_{t_{j+1}}$ based on original basis path set $B_{t_j}$. Here we emphasize $B'_{t_{j+1}}$ instead of original $B_{t_{j+1}}$ because $U_{t_{j+1}}$ may appear in different chains and paths in $B'_{t_{j+1}}$ can be discarded from multiple chains. For $U_0$, what we should do is to keep discarding paths from $B'_0$ based on $U_{t_{s_t}}$ for all $t$-th chain. The rule for Algorithm **SDVChain** to keep one path in $B'_{t_{j+1}}$ is that the shared layers with $B_{t_j}$ have the most concurrency and the unshared layers appear at least twice in $B'_{t_{j+1}}$ according to **Claim 1**. For $U_{t_1}$ in some $t$-th chain from the second group, enumerate all $t'$-th chain to find the last set $U_{t'_{i^*}} (t' \neq t)$ such that $p_{t'_{i^*}}$ shares layers with $p_{t_1}$ and discard the paths from $B'_{t_1}$ based on original $B_{t'_{i^*}}$. The aim of stretching path subdivision chain as long as possible is to reduce the enumeration as much as possible and the trick of path subdivision chain is that we only have to consider the path dependency between the consecutive sets in the chain.

To find the maximal independent substructure path set $P_{ind}^S$ from $P^S$ in Step 1, we borrow the concept of adjacent matrix in graph theory [15-17] and define adjacent matrix $M_r$ for path $p_r \in P^S$ as

$$M_r(j, l) = \begin{cases} 1, & \text{if there is edge from } j-\text{th layer to the } l-\text{th layer} \\ 0, & \text{otherwise} \end{cases} \quad \begin{matrix} j = 0,1, \ldots, L \\ l = 0,1, \ldots, L \end{matrix}$$

Reshape $M_r$ to substructure path vector $\alpha_r = [M_r(0,:), \ldots, M_r(j,:), \ldots, M_r(L,:)]$, so each $p_r$ can be expressed as a 0-1 substructure path vector $\alpha_r$ with $(L + 1)^2$ elements. It is easy to calculate the maximal independent path set $P_{ind}^S$ from $P^S$ by implementing linear algebra method, but it is infeasible to get basis path set $B$ in $P$ in this way, because $G^S$ is the simplest network with only one node in each layer and the size of regular network $G$ is too huge. To find the path subdivision and edge subdivision between two independent substructure paths, straightforward $L+1$-dimensional incident vector $\beta_r$ for substructure path $p_r$ is defined as

$$\beta_r(l) = \begin{cases} 1, & \text{if } p_r \text{ passes through the } l-th \text{ layer} \\ 0, & \text{otherwise} \end{cases}, \text{ where } l = 0,1,\dots,L.$$

For $p_r$ and $p_t$, let $X = \beta_r - \beta_t$. If $X$ contains 1 and -1, then $p_r$ and $p_t$ are not path subdivision to each other. If $X$ contains only 0 and -1, $p_t$ is path subdivision of $p_r$.

**Algorithm DEAH**
**Input:** Fully connected neural network $G = (V, E)$ with $L+1$ layers
**Output:** Path set $B$ of neural network $G$

**% Step 1. (The upper level) %**
Select randomly node $O_{i*}^l$ at the $l$-th ($l = 0, \dots, L$) layer of graph $G$. Set node subset $V^S = \{O_{i*}^0, \dots, O_{i*}^l, \dots, O_{i*}^L\}$ and edge subset $E^S = \{(O_{i*}^j, O_{i*}^l) \in E | O_{i*}^j \in V^S, O_{i*}^l \in V^S\}$. Let $P^S$ be the path set from $O_{i*}^0$ to $O_{i*}^L$ by breadth-first searching in $G^S = (V^S, E^S)$.
**For** each path $p_r \in P^S$ do
    Construct substructure path vector $\alpha_r$ and incident vector $\beta_r$.
**End For**
Calculate $R = Rank(\{\alpha_r\})$ by numerical linear algebra method.

**For** $r = 1: |P^S|$ do
    Let $U_r = \emptyset$. For each $t \in \{1, \dots, |P^S|\}$, let $X = \beta_r - \beta_t$. If $X$ contains only 0 and -1, set $U_r = U_r \cup \{t\}$.
**End For**
Pick up $p_{r_0} \in P^S$ with $U_{r_0} = \emptyset$. Sort $|U_{r_1}| \geq |U_{r_2}| \dots \geq |U_{r_{R-1}}| \dots \geq |U_{r_{|P^S|-1}}| \geq |U_{r_0}|$.
Set $i = 1$ and $A = \{\alpha_{r_0}\}$.
**While** $|A| < R$ do *% Rule out the possibility of Case 2 in the cause analysis of path dependency.*
    If $Rank(A \cup \{\alpha_{r_i}\}) = Rank(A)+1$ and $|U_{r_{i-1}}| \neq |U_{r_i}|$, let $A = A \cup \{\alpha_{r_i}\}$. If $Rank(A \cup \{\alpha_{r_i}\}) = Rank(A)+1$ but $|U_{r_{i-1}}| = |U_{r_i}|$, let $X = \beta_{r_{i-1}} - \beta_{r_i}$. If there is no 0 between any 1 and -1 in $X$ or no $U_{r_j}$ ($j < i-1$) such that $r_i, r_{i-1} \in U_{r_j}$, let $A = A \cup \{\alpha_{r_i}\}$.
    Let $i = i + 1$.
**End While**
Output $A$ as $\{U_r | r = 0,1,\dots,R-1\}$ with $U_0 = \emptyset$ and $|U_r|$ in descending order, $\{\beta_r\}$ and $P_{ind}^S = \{p_r\}$.

**%Step 2. (The lower level)** *% Get basis path set $B_r$ for $p_r$.*
**For** each $p_r \in P_{ind}^S$ do *% Induce sub-graph $G_r$ with the same structure as $p_r$.*
    Let $L_r = \{l | O_{i*}^l \in p_r\}$, $V_r = \{O^l \subset V | l \in L_r\}$ and $E_r = \{(O_i^{l_j}, O_{i'}^{l_{j+1}}) \in E | O_i^{l_j} \in O^{l_j}, O_{i'}^{l_{j+1}} \in O^{l_{j+1}}\}$. Set $G_r = (V_r, E_r)$. Call **Subroutine** ($G_r$) and output basis path set $B_r$.
**End For**

**%Step 3.** *% Eliminate path dependency from the union of basis paths.*
Set $X = \emptyset$ and let $t = 0$.
**For** $r = 1: R$ do *% If $U_r$ doesn't belong to any chain, start a new chain.*
    If $r \notin X$, let $t = t + 1$ and $U_{t_1} = U_r$. Find path subdivision chain $U_{t_1} \supset U_{t_2} \dots \supset U_{t_j} \dots \supset U_{t_{s_t}} \supset U_0$ by searching $\{U_r | r = r+1, \dots, R-1\}$ in order, till we couldn't stretch

the chain further. Let $Y_t = \{t_1, ..., t_j, ..., t_{s_t}\}$ and $X = X \cup Y_t$.
**End For**
Let $T = t$.

**For** $t = 1:T$ **do**
    Let $B'_0 = B_0$ and $B'_{t_j} = B_{t_j}$ for all $j$. % *Initialization for Algorithm SDVChain.*
    Let $Sh_t = \{r | E(p_{r_1}) \cap E(p_{t_1}) \neq \emptyset, r < t\}$ and $Q_t = \{t^*_j | $ the last $t^*_j$ satisfying $E(p_{t_1}) \cap E(p_{t^*_j}) \neq \emptyset, j \neq 1, t' \neq t\}$. Let $Q_t = Q_t \setminus Y_t$. % *One element may appear in multiple chains*
**End For**

**For** $t = 1:T$ **do**
    Call **SDVChain**( $\{B'_{t_j}\}_{j=1}^{S_t}, \{B_{t_j}\}_{j=1}^{S_t}, \{p_{t_j}\}_{j=1}^{S_t}, \{\beta_{t_j}\}_{j=1}^{S_t}, B'_0, p_0, \beta_0, G$), and output updated shrunk path set $\{B'_{t_j}\}_{j=2}^{S_t}$ and discarded path set $D'_0$. Let $B'_0 = B'_0 \setminus D'_0$.
    For each $k \in \{1, ..., |Q_t|\}$, call **SDVChain**($B'_{Q_t(k)}, B_{Q_t(k)}, p_{Q_t(k)}, \beta_{Q_t(k)}, B'_{t_1}, p_{t_1}, \beta_{t_1}, G$), output discarded path set $D_{t_1}$ and let $B'_{t_1} = B'_{t_1} \setminus D_{t_1}$. % $Q_t(k)$ *is the k-th element of* $Q_t$.
**End For**
Output path set $B = \cup_{t=1}^{T} (\cup_{j=1}^{S_t} B'_{t_j}) \cup B'_0$. ∎

**Theorem 1** Given fully connected neural network $G$ and $T$ path subdivision chains from Algorithm DEAH in $G$, i.e., $U_{t_1} \supset U_{t_2} ... \supset U_{t_j} ... \supset U_{t_{s_t}} \supset U_0$ with $t = 1, ..., T$. $B_{t_j}$ is the original basis path set of substructure path $p_{t_j}$ and $B'_{t_j}$ is the shrunk path set of $B_{t_j}$ after path discarding from Algorithm DEAH. Then the set $B = \cup_{t=1}^{T} (\cup_{j=1}^{S_t} B'_{t_j}) \cup B'_0$ is path independent and any path $p \in B_{t_j} \setminus B$ or $p \in B_0 \setminus B$ can be represented by $B$.
**Proof:** We now study the path independency between $p_k$ and $p_l$ from three cases.

**Case 1** $k = t_j$ $(j \neq 1)$, $l = t'_i$ $(i \neq 1)$ and $t \neq t'$. In this case $U_k$ and $U_l$ are not the first sets in the $t$-th and $t'$-th chain respectively. Let $B'_{t_j}$ be the shrunk path set of $B_{t_j}$ based on $B_{t_{j-1}}$ and $B'_{t'_i}$ be the shrunk path set of $B_{t'_i}$ based on $B_{t'_{i-1}}$ from the algorithm. We will prove that $B'_{t_j} \cup B'_{t'_i}$ from two different chains is path independent. Because of no path dependency within the same chain according to Lemma 1 (in Appendix B), assume that one path $p_{t_j,1}$ such that $p_{t_j,1} = p_{t'_i,1} - p_{t'_i,2} + p_{t_j,2}$, where $p_{t_j,1}, p_{t_j,2} \in B'_{t_j}$ and $p_{t'_i,1}, p_{t'_i,2} \in B'_{t'_i}$. Apparently, neither $U_{t'_i}$ nor $U_{t_j}$ could take $U_0$, otherwise $U_{t'_i}$ and $U_{t_j}$ are in the same chain. Let the layers $I'$ of substructure path $p_{t'_i}$ be the edge subdivision of the layers $I$ of $p_{t_j}$, as shown in Fig. 3(b). And the layers of $J$ in $p_{t_j}$ is the edge subdivision of $J'$ in $p_{t'_i}$. As for the structure uniqueness, paths $p_{t'_i,1}$ and $p_{t'_i,2}$ must have the same layers $I'$ and $J'$ to cancel each other, and paths $p_{t_j,1}$ and $p_{t_j,2}$ have the same layers $I$ and $J$. For the shared layers, $p_{t'_i,1}$ and $p_{t_j,1}$ must have $S'$ and $p_{t'_i,2}$ and $p_{t_j,2}$ have $S$ to swap between different substructures. No overlap between $I$ and $J$. Exchange edge set $J'$ and edge set $J$ in $p_{t'_i,1}$ and $p_{t_j,1}$ and exchange part $J'$ and part $J$ in $p_{t'_i}$ and $p_{t_j}$. Interestingly, we get two new substructure paths $p1$ and $p2$ (in Fig. 6). However, we notice that $p1, p_{t_j}$ and $p_{t'_i}$ (in Fig.3) are the path subdivisions of $p2$. It contradicts the way we select the maximal independent substructure path set.

**Case 2** $k = t_1$ and $l = t'_i$ $(i \neq 1)$. In this case, $U_k$ is the first set in the $t$-th and $U_l$ is inside the $t'$-th chain, where $t \neq t'$. $B'_{t'_i}$ is the shrunk path set of $B_{t'_i}$ based on $B_{t'_{i-1}}$ and $B'_{t_1}$ is shrunk from $B_{t_1}$. We prove that $B'_{t_1} \cup B'_{t'_i}$ is path independent. Lemma 2 (in Appendix

B) indicates that no path dependency would be produced for any $U_{t'_i}$ if we calculate $B'_{t_1}$ based on $p_{t'_{i^*}}$, which is the last substructure path in the chain $U_{t'_1} \supset U_{t'_2} \supset \cdots U_{t'_i} \supset \cdots U_{t'_{i^*-1}} \supset U_{t'_{i^*}} \cdots \supset U_{t'_{s_{t'}}}$ to have shared layers with $p_{t_1}$. If $i \leq i^*$, $B'_{t_1} \cup B'_{t'_i}$ is path independent from Lemma 2. If $i > i^*$, $B'_{t_1} \cup B'_{t'_i}$ is path independent, because $p_{t_1}$ and $p_{t'_i}$ don't share common edges.

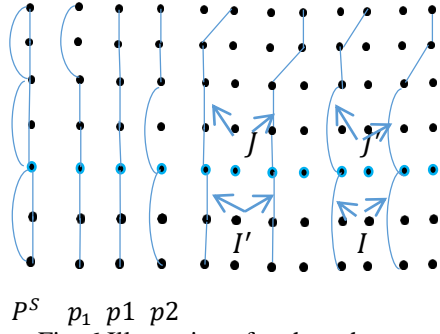

$P^S \quad p_1 \quad p1 \quad p2$
Fig. 6 Illustration of path exchange

**Case 3** $k = t_1$ and $l = t'_1$. In this case, both $U_k$ and $U_l$ are the first sets in their chains, where $t \neq t'$. In order to calculate $B'_{t_1}$, Algorithm **SDVChain** needs to find $p_{t'_{i^*}}$ in each $t'$-th chain and discard the corresponding paths from $B'_{t_1}$ based on $B_{t'_{i^*}}$. This procedure guarantees the path independency for $B'_{t_1} \cup B'_{t'_1}$, according to Lemma 2.

According to Algorithm **SDVChain,** $D'_0$ is the path set which will be discarded from $B'_0$, based on original $B_{r_{s_t}}$ in the $t$-th chain. $B'_0$ is initialized as original $B_0$ and is updated as $B'_0 = B'_0 \setminus D'_0$ iteratively for $t = 1 \ldots T$. And any $p \in B_0 \setminus B'_0$ can be represented by $B'_0 \cup B_{r_{s_t}}$ for some $t$. Since $U_{t_1} \supset U_{t_2} \ldots \supset U_{t_j} \ldots \supset U_{t_{s_t}} \supset U_0$ is path subdivision chain, so $B'_0 \cup B_{r_j}$ is path independent for $j = 1, \ldots, s_t$ in the $t$-th chain. Therefore, $B = \cup_{t=1}^T (\cup_{j=1}^{s_t} B'_{t_j}) \cup B'_0$ is path independent based upon Claim 1 and Claim 2.

Note set $U_{t_j}(j \neq 1)$ can appear in multiple chains but the first set $U_{t_1}$ can only appear in exactly $t$-th chain. Lemma 1 proves that any path $p \in B_{t_j} \setminus B'_{t_j}$ can be represented by $B_{t_{j-1}} \cup B'_{t_j}$ recursively till $B_{t_1}$. If $U_{t_j}$ belongs to the $t'$-th chain and the $t$-th chain, i.e., $U_{t'_1} \supset \cdots U_{t'_i} \supset U_{t_j}, p \in B_{t_j} \setminus B'_{t_j}$ can be represented either by $B_{t_{j-1}} \cup B'_{t_j}$ or by $B_{t'_i} \cup B'_{t_j}$. Hence, any path $p \in B_{t_j} \setminus B'_{t_j} (j \neq 1)$ can be represented by $B$. Moreover, any path $p \in B_{t_1} \setminus B'_{t_1}$ can be represented by $B$, because Lemma 2 concludes $p$ can be represented by some $B_{t'_{i^*}} \cup B'_{t_1}$ and any path in $B_{t'_{i^*}}$ can be represented by $B$. Furthermore, any $p \in B_0 \setminus B'_0$ can be represented by $B$ for $p$ can be represented by $B'_0 \cup B_{r_{s_t}}$ for some $t$. ∎

**Theorem 2** Given fully connected neural network $G$ and independent path set $B = \cup_{t=1}^T (\cup_{j=1}^{s_t} B'_{t_j}) \cup B'_0$ output from Algorithm DEAH, where $T$ is the number of path subdivision chains and $B'_{t_j}$ is the shrunk path set of $B_{t_j}$ in $G_{t_j}$. Then any path $p \in P$ from the input layer to the output layer in $G$ can be represented by $B$.

**Proof:** Any path $p \in P$ can be represented in the hierarchical way, first at substructure level and then at basis path level. If the structure of $p$ is $p_r \in P^S_{ind}$ but $p \notin B_r$, it is trivia that path $p$ can be represented by $B_r$. If the structure of $p$ is out of the structure range of $P^S_{ind}$, the structure of $p$ can be expressed as $p_{r_1} \ldots + p_{r_j} \ldots + p_{r_d} - p_{s_1} \ldots - p_{s_h} \ldots - p_{s_m}$, where $r_1, \ldots, r_d, s_1, \ldots, s_m$ are distinct, $p_{r_j} \in P^S_{ind}$ and $p_{s_h} \in P^S_{ind}$ ($1 \leq j \leq d$, $1 \leq h \leq m$). Suppose the randomly node set is $V^S = \{O^0_{i^*}, \ldots, O^l_{i^*}, \ldots, O^L_{i^*}\}$ and target path $p$ passes

$\{O_{i'}^0, ..., O_{i'}^l, ..., O_{i'}^L\}$ sequentially. Define new node set $V' = \{O_{i''}^0, ..., O_{i''}^l, ..., O_{i''}^L\}$, where $l = 0, 1, ..., L$. If $p$ skips over the $l$-th layer, set $O_{i''}^l = O_{i^*}^l$. If $p$ passes through the $l$-th layer and $O_{i'}^l \neq O_{i^*}^l$, set $O_{i''}^l = O_{i'}^l$. For each $p_{r_j}$ and $p_{s_h}$, construct new path $p'_{r_j}$ and $p'_{s_h}$ passing through node set $V'$ but $p'_{r_j}$ and $p'_{s_h}$ keep the same structure as $p_{r_j}$ and $p_{s_h}$ under $V^S$. $V'$ covers all nodes of path $p$, and node $O_{i^*}^l \in V^S$ corresponds to $O_{i''}^l \in V'$. In original graph $G$, path $p$ therefore can be accordingly expressed as $p = p'_{r_1} ... + p'_{r_j} ... + p'_{r_d} - p'_{s_1} ... - p'_{s_h} ... - p'_{s_m}$. Moreover, new path $p'_{r_j}$ can be represented by basis path set $B_{r_j}$ and $p'_{s_h}$ can be represented by $B_{s_h}$. Hence, $p$ can be represented by $B = \cup_{t=1}^{T} (\cup_{j=1}^{s_t} B'_{t_j}) \cup B'_0$ according to Theorem 1. Therefore, $B$ is a basis path set for neural network $G$. ∎

Theorem 1 and Theorem 2 prove that Algorithm DEAH can find the basis path set $B$ for *Basis Path Set Searching* problem in regular graph $G$. Next, we will prove that Algorithm DEAH can be completed in polynomial time.

**Theorem 3** The time complexity of **Algorithm DEAH** to solve *Basis Path Set Searching* problem in regular graph $G$ is $\mathcal{O}(RL^2W_{max}^2) + \mathcal{O}\left(\max(R, T^2) \cdot (L + B_{max}^3)\right)$, where $R = |P_{ind}^S|$, $W_{max} = \max_{0 \leq l \leq L}\{|O^l|\}$ and $B_{max} = \max_{r \in \{1, 2, ..., R\}} |B_r|$.

**Proof:** There are three major steps in Algorithm DEAH.
Step 1 of Algorithm DEAH finds $V^S$ in $\mathcal{O}(L)$ time, and searches $E^S$ in $\mathcal{O}(m)$ time and $P^S$ in at most $\mathcal{O}(m)$ time. In order to get $P_{ind}^S$, we compute all $U_r$ in at most $\mathcal{O}(m^2)$ time and compute $Rank(A)$ by searching $\{U_{r_1}, U_{r_2}, ..., U_{r_{|P^S|-1}}\}$ in at most $\mathcal{O}(m)$ time, since $|P^S| \leq m$. So, Step 1 runs in $\mathcal{O}(3m + m^2 + L) = \mathcal{O}(m^2)$ time.

Step 2 calls **Subroutine** $(G_r)$ to find basis path set $B_r$ of substructure $p_r (r = 1, 2, ..., R)$. **Lemma 3** (in Appendix C) proves that **Algorithm Subroutine**$(G)$ runs in time $\mathcal{O}(L^2W_{max}^2)$ in fully connected graph $G$ without edge-skipping. Let $W_{max} = \max_{0 \leq l \leq L}\{|O^l|\}$ in network $G$. Therefore, the time complexity of Step 2 is $\mathcal{O}(RL^2W_{max}^2)$.

In Step 3, there are two parts for the computation of shrunk path sets. One part is to compute $B'_{t_{j+1}} (j \neq 0)$. In every $t$-th chain we call **SDVChain** to compute $B'_{t_{j+1}}$ based on $B_{t_j}$ iteratively. It takes $\mathcal{O}(L)$ time to find common layers between $p_{t_j}$ and $p_{t_{j+1}}$, $\mathcal{O}\left(|B_{t_j}|\right)$ time to separate the shared layers and unshared layers in $B_{t_j}$ and $\mathcal{O}\left(|B_{t_{j+1}}|\right)$ time to separate the layers of $B_{t_{j+1}}$. To get unique unshared layers $\{Ep_{t_j,i}\}$ needs $\mathcal{O}\left(|B_{t_j}|^2\right)$ time. For each $Ep_{t_j,i}$, it needs at most $\mathcal{O}\left(|B_{t_j}|\right)$ time to compute $UCP_i$ and $UCP_i^*$ in $B_r$ and $\mathcal{O}\left(|B_{t_j}||B_{t_{j+1}}|\right)$ time to get unshared layers set $IEp_i$ in $B'_{r+1}$. Since there are at most $\mathcal{O}\left(|B_{t_j}|\right)$ elements in $\{Ep_{t_j,i}\}$, so this phase needs at most $\mathcal{O}\left(|B_{t_j}| \cdot (|B_{t_j}| + |B_{t_j}||B_{t_{j+1}}|)\right) = \mathcal{O}(B_{max}^3)$ time, where $B_{max} = \max_{r \in \{1, 2, ..., R\}} |B_r|$. Furthermore, there are $T$ chains, every $t$-th chain needs to call **SDVChain** for $s_t$ times, some $U_{t_{j+1}}$ may appear in most $T$ chains and the path set after $U_{t_{j+1}}$ in the chain only needs computing once. Hence, we need at most $\mathcal{O}(R + T) \cdot \mathcal{O}\left(L + |B_{t_j}| + |B_{t_{j+1}}| + |B_{t_j}|^2 + B_{max}^3\right) = \mathcal{O}\left((R + T)(L + B_{max}^3)\right) = \mathcal{O}(R(L + B_{max}^3))$ time. In the other part of Step 3, every $U_{t_1}$ needs to enumerate all elements in $Q_t$ to call **SDVChain** to update $B'_{t_1}$ based on $B_{Q_t(k)}$ iteratively. This simple version of **SDVChain** would take at most $\mathcal{O}(L + B_{max}^3)$ time for each iteration and there are at most $T$ elements in $B_{Q_t(k)}$, i.e., totally $\mathcal{O}(T(L + B_{max}^3))$ time. Because there are $T$ chains, so this phase takes $\mathcal{O}(T^2(L + B_{max}^3))$ time. Therefore, Step 3 takes totally $\mathcal{O}(R(L + B_{max}^3)) + \mathcal{O}(T^2(L + B_{max}^3)) = \mathcal{O}(\max(R, T^2)(L + B_{max}^3))$ time.

In sum, the total time complexity of **Algorithm DEAH** is $\mathcal{O}(m^2) + \mathcal{O}(RL^2W_{max}^2) + \mathcal{O}(\max(R, T^2)(L + B_{max}^3)) = \mathcal{O}(RL^2W_{max}^2) + \mathcal{O}(\max(R, T^2)(L + B_{max}^3))$. ∎

It is obvious that the computation complexity of **Algorithm DEAH** to solve *Basis Path Set Searching* problem depends heavily on the structure of the network, *i.e.*, the maximal layer width and edge-skipping over layers. Usually $B_0$ is the basis path set taking $B_{max}$. Each $p_r \in P_{ind}^S$ represents one type of substructure information about edge-skipping over layers in network $G$, and different types of substructures can be combined in a variety of ways. Thus, there is no constraint about the graph structure levying on our Algorithm DEAH, which breaks the bottleneck of the algorithm in [14] and generalizes it to more practical networks. Though Algorithm DEAH considers only one underlying structure path, but it can be easily extended to the network with multiple underlying structure paths.

## 5 Conclusion

In regular fully connected network $G$, the shared layers between two independent substructure paths bring up the combinatoric possibility of path dependency when combining the basis path sets from these substructures. Algorithm DEAH is designed to eliminate such path dependency and the trick of effective elimination is the path subdivision chain. The theoretical proofs guarantee the feasibility of Algorithm DEAH for *Basis Path Set Searching* problem. The paper generalizes the specific network structure with equal layer and without edge-skipping to more practical network and provides one methodology to solve *Basis Path Set Searching* problem in more general neural network. This work can help facilitate the theoretic research and applications of $\mathcal{G}$-SGD algorithm in more practical scenarios.

## Appendix A

Some definitions may be referred.

**Definition 10** (path addition to a graph)[14] Given a graph $H$ and a path $p$, we denote the path addition by $H + p$ with $V(H + p) = V(H) \cup V(p)$ and $E(H + p)$ being the disjoint union of $E(G)$ and $E(p)$ (Parallel edges may arise).

**Definition 11** (path removal from a graph)[14] Given a graph $H$ and one path $p \subseteq E(H)$, the removal of the path $p$ from the graph $H$ is defined as $H - p$ with $E(H - p) = E(H) \setminus E(p)$ and $V(H - p) = V(H) \setminus \{v \in V(H) | v \text{ is an isolated vertex after } E(H) \setminus E(p)\}$.

**Definition 12** (structure path)[14] Given fully connected neural network $G$, if all paths in $P$ from the input layer to the output layer passes through the same layers consecutively, any path $p \in P$ can be called as the structure path of neural network $G$, since it can express the structure information of $G$.

**Appendix B**
Some properties and theoretical proofs related to path subdivision chains are given in this section. Suppose $T$ path subdivision chains $\{U_{t_1} \supset U_{t_2} ... \supset U_{t_j} ... \supset U_{t_{s_t}} \supset U_0 | t = 1, ..., T\}$ are output from Algorithm DEAH.

**Lemma 1** Given path subdivision chain $U_{t_1} \supset U_{t_2} ... \supset U_{t_j} ... \supset U_{t_{s_t}} \supset U_0$ in fully connected neural network $G$, where path subdivision set $U_{t_j}$ and basis path set $B_{t_j}$ correspond to maximal independent substructure path $p_{t_j}$. Especially $U_0$ and $B_0$ correspond to the underlying substructure path $p_0$. Then, $B_{t_1} \cup B'_{t_2} \cup ... B'_{t_j} ... \cup B'_{t_{s_t}}$ is path independent, where $B'_{t_j}$ output from Algorithm SDVChain is the shrunk path set of $B_{t_j}$ based on original basis path set $B_{t_{j-1}}$ for $j = 2,3, ..., s_t$.

**Proof:** We use induction on the $j$-th subdivided substructure path to prove that $B_{t_1} \cup B'_{t_2} \cup ... B'_{t_j} ... \cup B'_{t_{s_t}}$ is path independent. The inputs of Algorithm **SDVChain** are $\{B'_{t_j}\}_{j=1}^{s_t}, \{B_{t_j}\}_{j=1}^{s_t}$, $\{p_{t_j}\}_{j=1}^{s_t}, \{\beta_{t_j}\}_{j=1}^{s_t}$, $B'_0$, $p_0$ and $\beta_0$, where $B'_{t_j} = B_{t_j}$ and $B'_0 = B_0$ for algorithm initiation.

Basis step: $j = 1$ and $U_{t_1} \supset U_{t_2}$. Step 1 of Algorithm **SDVChain** picks up $UCP_j^*$ with the most frequent occurrence in $UCP_j$ for each unique $Ep_{t_1,j}$ of $B_{t_1}$. The trick is that the edge between every two common layers with the most frequency must be the direct path according to the construction rule [14] of basis path. Step 2 of the algorithm searches the unshared layers set $IEp_j$ in $B_{t_2}$ which has the shared layers from $UCP_j$ and discard the element with frequency less than 2 from $IEp_j$. According to Claim 1, the paths with multiple repetition at unshared layers in $IEp_j$ can cause path dependency. Next, the algorithm outputs $B'_{t_2}$ by discarding the paths from $B_{t_2}$ whose unshared layers are from $IEp_j$ element and shared layers are not $UCP_j^*$. Hence, any path $p \in B_{t_1} \cup B'_{t_2}$ couldn't be represented by $B_{t_1} \cup B'_{t_2} \setminus \{p\}$ and any path $p_{t_2 u'} \in B_{t_2} \setminus B'_{t_2}$ can be represented by $B_{t_1} \cup B'_{t_2}$. So, $B_{t_1} \cup B'_{t_2}$ is path independent for the basic step.

Induction step: $j \geq 2$. Suppose that $B_{t_1} \cup B'_{t_2} \cup ... B'_{t_j}$ is an independent path set and any path $p \in B_{t_{j-1}} \setminus B'_{t_j}$ can be represented by $B_{t_{j-1}} \cup B'_{t_j} \setminus \{p\}$. Let $B'_{t_{j+1}}$ be the shrunk path set of $B_{t_{j+1}}$ based on original $B_{t_j}$ from Algorithm. We will prove that any path $p \in B_{t_1} \cup B'_{t_2} \cup ... B'_{t_j} \cup B'_{t_{j+1}}$ couldn't be represented by $B_{t_1} \cup B'_{t_2} \cup ... B'_{t_j} \cup B'_{t_{j+1}} \setminus \{p\}$. Assume path $p' \in B_{t_1} \cup B'_{t_2} \cup ... B'_{t_j} \cup B'_{t_{j+1}}$ can be represented by $B_{t_1} \cup B'_{t_2} \cup ... B'_{t_j} \cup B'_{t_{j+1}} \setminus \{p'\}$. According to Claim 2, consider the simplest form that $p' = p_{t_{j+1},1} - p_{t_{j+1},2} + p''$, where $p_{t_{j+1},1} \in B'_{t_{j+1}}$ and $p_{t_{j+1},2} \in B'_{t_{j+1}}$ such that $p' \in B'_{t_m}$ and $p'' \in B'_{t_m}$ with $m \leq j - 1$. Similar to basic step, any path $p \in B_{t_j} \setminus B'_{t_{j+1}}$ can be represented by $B_{t_j} \cup B'_{t_{j+1}} \setminus \{p\}$, so it is obvious that $p', p'' \notin B'_{t_{j+1}}$ and $p', p'' \notin B'_{t_j}$.

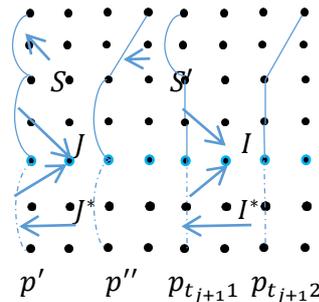

Fig. 7 Edge operation in subdivided unshared layers

Among the specific chain $U_{t_m} \supset U_{t_j} \supset U_{t_{j+1}}$, substructure path $p_{t_{j+1}}$ is the path subdivision of $p_{t_j}$, and $p_{t_j}$ is the path subdivision of $p_{t_m}$. So $p_{t_{j+1}}$ is the most subdivided path. As shown in Fig. 7, $p_{t_{j+1},1}$ and $p_{t_{j+1},2}$ must have same sub-path $I$ and paths $p'$ and $p''$ must have $J$ to cancel the unique sub-paths, since $p_{t_{j+1}}$ is the path subdivision of $p_{t_m}$ at the layers of $J$. For the shared layers of $p_{t_m}$ and $p_{t_{j+1}}$, $p'$ and $p_{t_{j+1}1}$ share sub-path $S$ and $p''$ and $p_{t_{j+1}2}$ share $S'$. $p_{t_j}$ should be the path subdivision of $p_{t_m}$ in some part of $J$, i.e., the dashed lowest part $J^*$ of $J$. Hence, the dashed part $I^*$ in $J^*$ of $p_{t_j}$ subdivided by $p_{t_{j+1},1}$ and $p_{t_{j+1},2}$ are the same. However, the shared layers such as $I - I^* + S$ in $p_{t_{j+1}1}$ and $I - I^* + S'$ in $p_{t_{j+1}2}$ are different as shown in Fig. 7. This contradicts the Algorithm **SDVChain** which keeps only the most frequent shared layers from $B_{t_j}$ to $B'_{t_{j+1}}$.

Repeat this induction step till $j = s_t$. ∎

The trick of Lemma 1 is that we can construct path subdivision chain to avoid the enumeration within the chain. Furthermore, if $U_{t_j}$ belongs to more than one chain, i.e., $U_{t'_i} \supset U_{t_j}$ in the $t'$-th chain and $U_{t_j}$ is the nearest path subdivision of $U_{t'_i}$, Algorithm **SDVChain** will shrink $B'_{t_j}$ iteratively by discarding the paths from $B'_{t_j}$ based on $B'_{t'_i}$. In this case, it is trivia that path set $B_{t_1} \cup B'_{t_2} \cup \ldots B'_{t_j} \ldots \cup B'_{t_{s_t}}$ is an independent path set and any path $p \in B_{t_j} \setminus B'_{t_j}$ can be represented by either $B_{t_{j-1}} \cup B'_{t_j}$ or $B_{t'_i} \cup B'_{t_j}$.

**Lemma 2** Given the first set $U_{t_1}$ in the $t$-th chain and some $i^* \in Q_t$, where $p_{t'_{i^*}}$ is the last substructure path in the $t'$-th chain to have shared layers with $p_{t_1}$ ($t' \neq t$ and $t'_{i^*} \neq 0$). $B'_{t_1}$ is the shrunk path set output from Algorithm **SDVChain** by deleting the paths from $B_{t_1}$ based on all $B_{t'_i}$ iteratively. Then $B'_{t_1} \cup B'_{t'_i}$ is path independent for any $i \in \{1, \ldots, i^* - 1\}$ in the $t'$-th chain.

**Proof:** Suppose the $t'$-th path subdivision chain be $U_{t'_1} \supset U_{t'_2} \supset \cdots U_{t'_i} \supset \cdots U_{t'_{i^*-1}} \supset U_{t'_{i^*}} \ldots \supset U_{t'_{s_{t'}}}$. $Q_t$ would rules out some substructure paths: 1) substructure path after $p_{t'_{i^*}}$ because it doesn't share layers with $p_{t_1}$. 2) the sets from the $t$-th chain because of Lemma 1. 3) $p_{r_1}(r < t)$ which shares edges with $p_{t_1}$, because the algorithm only run one direction. Assume there exist two paths $p_{t'_i,1}, p_{t'_i,2} \in B'_{t'_i}$ with $i < i^*$ and two paths $p_{t_1,1}, p_{t_1,2} \in B'_{t_1}$ such that $p_{t'_i,1} - p_{t'_i,2} = p_{t_1,1} - p_{t_1,2}$. Since Step 2 of Algorithm DEAH demands the substructure path be subdivided in the unshared layers by the next substructure path when forming the $t'$-th chain, there must be corresponding $p_{t'_{i^*},1}$ and $p_{t'_{i^*},2}$ in $B_{t'_{i^*}}$ such that $p_{t'_{i^*},1} - p_{t'_{i^*},2} = p_{t_1,1} - p_{t_1,2}$ for $p_{t'_{i^*}}$ shares layers with $p_{t'_i}$. But, Algorithm SDVChain discards $p_{t_1,1}$ or $p_{t_1,2}$ based on original $B_{t'_{i^*}}$ already. This contradicts the assumption. So $B'_{t_1} \cup B'_{t'_i}$ is path independent. On the other hand, any path $p \in B_{t_1} \setminus B'_{t_1}$ can be represented by $B'_{t_1}$ and some $B_{t'_{i^*}}$. And any path $p \in B_{t'_i}$ can be represented by $B'_{t'_i} \cup B'_{t'_{i^*-1}}$. ∎

Lemma 2 indicates that no path dependent would be produced for any $U_{t'_i}$ if we calculate $B'_{t_1}$ based on $B_{t'_{i^*}}$, though some $p_{t'_i}$ ($i < i^*$) shares common layers with $p_{t'_{i^*}}$. The second trick of substructure path subdivision chain is to avoid enumerating the full chain for the first set of another chain.

**Appendix C**
**Lemma 3** The time complexity of **Algorithm Subroutine**$(G)$ in fully connected graph $G$

without edge skipping is $\mathcal{O}(L^2 W_{max}^2)$, where $W_{max} = \max\limits_{0 \leq l \leq L}\{|O^l|\}$.

**Proof:** Algorithm **Subroutine** takes at most totally $\mathcal{O}(W_{max}^2 L)$ time to construct all sub-graph $G(k)$ ($k = 0: L-1$). It takes $\mathcal{O}(W_{max}^2)$ time to search direct path set $P_{dir}^{(k)}$ and cross path set $P_{cross}^{(k)}$ in $G(k)$, so total running time for all $G(k)$ is $\mathcal{O}(W_{max}^2 L)$. When updating $P_{dir}^{(k)}$ and $P_{cross}^{(k)}$, the average number of paths from the lower layer is at most $\mathcal{O}(kW_{max})$ for each node $O_i^k$ and there are at most $W_{max}$ nodes for $k$-th layer, so the total time complexity for updating is $\mathcal{O}(\frac{L(L-1)}{2} W_{max}^2)$. The time for classifying paths to $k+1$-th layer is the same as for updating $P_{dir}^{(k)}$ and $P_{cross}^{(k)}$. So the total running time for Algorithm Subroutine is $\mathcal{O}(W_{max}^2 L + W_{max}^2 L + 2 \times \frac{L(L-1)}{2} W_{max}^2) = \mathcal{O}(L^2 W_{max}^2)$. ∎

**SDVChain($\{B_r'\}_{r=1}^s, \{B_r\}_{r=1}^s, \{p_r\}_{r=1}^s, \{\beta_r\}_{r=1}^s, B_0, p_0, \beta_0, G$):**
**Input:** Shrunk path set $B_r'$, basis path set $B_r$ and $\beta_r$ regarding $p_r$, independent path set $B_0$ and $\beta_0$ regarding $p_0$ in fully connected network $G$.
% $p_r$ is subdivided by $p_{r+1}$ and $p_s$ is subdivided by $p_0$
**Output:** Updated shrunk path set $\{B_r'\}_{r=2}^s$ and discarded path set $D_0'$.

Let $B_{s+1}' = B_0, p_{s+1} = p_0$ and $\beta_{s+1} = \beta_0$.
**For** $r = 1: s$ **do**
    Search shared layers $\{(O^{l_k'}, O^{l_k''})\}_{k=1}^K$ for $p_r$ and $p_{r+1}$, according to $\beta_r$ and $\beta_{r+1}$. Find unique unshared layers $\{Ep_{r,j}\}$ in $B_r$. Let $DiscardPath = \emptyset$.
    **For** each unique $Ep_{r,j}$ **do**
        Find shared layers set $UCP_j$ in $B_r$ with $Ep_{r,j}$ as the unshared layers. Calculate $UCP_j^*$ with the most frequent occurrence for each element in $UCP_j$. % Find the most frequent edge set in the shared layers.
        Construct the unshared layers set $IEp_j$ in $B_{r+1}'$ which has the shared layers from $UCP_j$. Discard the element with frequency less than 2 from $IEp_j$. % Get rid of the path whose unshared layers appears only once, which couldn't cause path dependency.
        Let $DiscardPath = DiscardPath \cup \{p_{r+1,i} \in B_{r+1}' |$ the unshared layers of $p_{r+1,i}$ is from $IEp_j$ and the shared layers are not $UCP_j^*$, $i \in \{1, \ldots, |IEp_j|\}\}$. % Discard the path which brings up dependency when combining $B_r$ with $B_{r+1}'$.
    **End For**
    Update $B_{r+1}' = B_{r+1}' \setminus DiscardPath$ and set $D_{r+1}' = DiscardPath$.
**End For**
Set $D_0' = D_{s+1}'$.

**Subroutine($G$):**
**Input:** Fully connected neural network $G = (V, E)$ without any edge-skipping over layers.
**Output:** Basis path set $B$ in graph $G$
**For** $k = 0: L-1$ **do**
    Let $E^k = \{e \in G | e$ leaves from $k$-th layer and enters $k+1$-th layer $\}$.
    % **Step 1**. Construct the direct path set.
    Let sub-graph $G(k) = (O^k \cup O^{k+1}, E^k)$.
    **If** $|O^k| \geq |O^{k+1}|$ **do**
        Find $|O^{k+1}|$ direct vertex disjoint paths by depth-first searching, and let the direct path set be $P_{dir}^{(k)}$.
        **For** $v \in O^k \setminus V(P_{dir}^{(k)})$ **do**
            Pick up one node $O_{i'}^{k+1} \in O^{k+1}$ randomly and construct path $(v, O_{i'}^{k+1})$. Set $P_{dir}^{(k)} = P_{dir}^{(k)} \cup (v, O_{i'}^{k+1})$.

    **End For**
   **Else do**
    Find $|O^k|$ direct vertex disjoint paths by depth-first searching, and let the direct path set be $P_{dir}^{(k)}$.
   **End If**
   **For** $i = 1, 2, \ldots, |O^k|$ **do**
    Let the path set $P_{dir}(O_i^k) = \{ p \in P_{dir}^{(k)} | \text{the tail of } p \text{ is node } O_i^k \}$.
   **End For**
   *% Step 2. Construct the cross path.*
   Set cross path set $P_{cross}^{(k)} = E^k \setminus E(P_{dir}^{(k)})$.
   **For** $i = 1, 2, \ldots, |O^k|$
    Let the path set $P_{cross}(O_i^k) = \{p \in P_{cross}^{(k)} | \text{the tail of } p \text{ is node } O_i^k\}$.
   **End For**
   *% Step 3. Concatenate the direct paths and cross paths from the $k - 1$-th layer.*
   **If** $k \neq 0$  *% If $k = 0$, there is no concatenation for any path and go to Step 4 directly.*
    **For** $i = 1, 2, \ldots, |O^k|$ **do** *% Form $|P(O_i^k)|$ direct paths and extend all cross paths.*
     Let $P_{dir}(O_i^k) = \{p_0 + p_1 | p_1 \in P_{dir}(O_i^k), p_0 \in P(O_i^k)\}$ for node $O_i^k \in O^k$. Select one path $p^* \in P(O_i^k)$ randomly and let $P_{cross}(O_i^k) = \{ p^* + p_1 | p_1 \in P_{cross}(O_i^k)\}$.
    **End For**
    Update direct path set $P_{dir}^{(k)} = \cup_{O_i^k \in O^k} P_{dir}(O_i^k)$ and cross path set $P_{cross}^{(k)} = \cup_{O_i^k \in O^k} P_{cross}(O_i^k)$.
   **End If**
   *% Step 4. Classify the paths for the nodes in the $k + 1$-th layer.*
   **For** $i = 1, 2, \ldots, |O^{k+1}|$ **do**
    Set the path set $P(O_i^{k+1}) = \{p \in P_{dir}^{(k)} \cup P_{cross}^{(k)} | \text{ the head of } p \text{ is node } O_i^{k+1}\}$.
   **End For**
  **End for**
  Output basis path set $B = P_{dir}^{(L-1)} \cup P_{cross}^{(L-1)}$.   ∎